\documentclass[prb,twocolumn,aps,nobibnotes,amsmath,amssymb,superscriptaddress,floatfix]{revtex4-2}

\UseRawInputEncoding
\usepackage{graphicx}% Include figure files
\usepackage{dcolumn}% Align table columns on decimal point
\usepackage{bm}% bold math
\usepackage{here} 
\usepackage{epsf}
\usepackage{epstopdf}
\usepackage{color}
\usepackage{upgreek}
\usepackage{braket}
\usepackage{tabularx}

\begin{document}

%\title{Electronic structures and ground-state $4f$ symmetry on CeNi$_{2}$Ge$_{2}$\\ probed by advanced electron spectroscopies}
\title{Impact of the ground-state $4f$ symmetry for anisotropic $cf$-hybridization \\in the heavy fermion superconductor CeNi$_{2}$Ge$_{2}$}
%\\ probed by advanced electron spectroscopies}

%----------------------------------------------- Author ---------------------------------------------------------

\author{H. Fujiwara}
\email{fujiwara@mp.es.osaka-u.ac.jp}
\affiliation{Division of Materials Physics, Graduate School of Engineering Science, Osaka University, Toyonaka, Osaka 560-8531, Japan}
\affiliation{RIKEN SPring-8 Center, Sayo, Hyogo 679-5148, Japan}
\affiliation{Spintronics Research Network Division, Institute for Open and Transdisciplinary Research Initiatives,Osaka University, Yamadaoka 2-1, Suita, Osaka, 565-0871, Japan}

\author{Y. Nakatani}
\affiliation{Division of Materials Physics, Graduate School of Engineering Science, Osaka University, Toyonaka, Osaka 560-8531, Japan}
\affiliation{RIKEN SPring-8 Center, Sayo, Hyogo 679-5148, Japan}

\author{H. Aratani}
\affiliation{Division of Materials Physics, Graduate School of Engineering Science, Osaka University, Toyonaka, Osaka 560-8531, Japan}
\affiliation{RIKEN SPring-8 Center, Sayo, Hyogo 679-5148, Japan}

\author{Y. Kanai-Nakata}
\affiliation{Division of Materials Physics, Graduate School of Engineering Science, Osaka University, Toyonaka, Osaka 560-8531, Japan}
\affiliation{RIKEN SPring-8 Center, Sayo, Hyogo 679-5148, Japan}
\affiliation{Department of Physical Sciences, Ritsumeikan University, Kusatsu, Shiga 525-8577, Japan}

\author{K. Yamagami}
\affiliation{Division of Materials Physics, Graduate School of Engineering Science, Osaka University, Toyonaka, Osaka 560-8531, Japan}
\affiliation{RIKEN SPring-8 Center, Sayo, Hyogo 679-5148, Japan}

\author{S. Hamamoto}
\affiliation{Division of Materials Physics, Graduate School of Engineering Science, Osaka University, Toyonaka, Osaka 560-8531, Japan}
\affiliation{RIKEN SPring-8 Center, Sayo, Hyogo 679-5148, Japan}

\author{\\T. Kiss}
\affiliation{Division of Materials Physics, Graduate School of Engineering Science, Osaka University, Toyonaka, Osaka 560-8531, Japan}
\affiliation{RIKEN SPring-8 Center, Sayo, Hyogo 679-5148, Japan}

\author{A. Yamasaki}
\affiliation{RIKEN SPring-8 Center, Sayo, Hyogo 679-5148, Japan}
\affiliation{Faculty of Science and Engineering, Konan University, Kobe 658-8501, Japan}

\author{A. Higashiya}
\affiliation{RIKEN SPring-8 Center, Sayo, Hyogo 679-5148, Japan}
\affiliation{Faculty of Science and Engineering, Setsunan University, Neyagawa, Osaka 572-8508, Japan}

\author{S. Imada}
\affiliation{RIKEN SPring-8 Center, Sayo, Hyogo 679-5148, Japan}
\affiliation{Department of Physical Sciences, Ritsumeikan University, Kusatsu, Shiga 525-8577, Japan}

\author{A. Tanaka}
\affiliation{Department of Quantum Matter, ADSM, Hiroshima University, Higashi-hiroshima, Hiroshima 739-8530, Japan}

\author{K. Tamasaku}
\affiliation{RIKEN SPring-8 Center, Sayo, Hyogo 679-5148, Japan}

\author{M. Yabashi}
\affiliation{RIKEN SPring-8 Center, Sayo, Hyogo 679-5148, Japan}

\author{T. Ishikawa}
\affiliation{RIKEN SPring-8 Center, Sayo, Hyogo 679-5148, Japan}

\author{\\A. Yasui}
\affiliation{Materials Sciences Research Center, Japan Atomic Energy Agency, Sayo, Hyogo 679-5148, Japan}
\affiliation{Japan Synchrotron Radiation Research Institute, Sayo, Hyogo 679-5198, Japan}

\author{H. Yamagami}
\affiliation{Materials Sciences Research Center, Japan Atomic Energy Agency, Sayo, Hyogo 679-5148, Japan}
\affiliation{Faculty of Science, Kyoto Sangyo University, Kyoto 603-8047, Japan}

\author{J. Miyawaki}
\affiliation{The Institute for Solid State Physics (ISSP), The University of Tokyo, Kashiwa, Chiba 277-8581, Japan}

\author{A. Miyake}
\affiliation{The Institute for Solid State Physics (ISSP), The University of Tokyo, Kashiwa, Chiba 277-8581, Japan}

%\author{M. Tokunaga}
%\affiliation{The Institute for Solid State Physics (ISSP), The University of Tokyo, Kashiwa, Chiba 277-8581, Japan}

\author{\\T. Ebihara}
\affiliation{Department of Physics, Shizuoka University, Shizuoka 422-8529, Japan}

\author{Y. Saitoh}
\affiliation{Materials Sciences Research Center, Japan Atomic Energy Agency, Sayo, Hyogo 679-5148, Japan}

\author{A. Sekiyama}
\affiliation{Division of Materials Physics, Graduate School of Engineering Science, Osaka University, Toyonaka, Osaka 560-8531, Japan}
\affiliation{RIKEN SPring-8 Center, Sayo, Hyogo 679-5148, Japan}
\affiliation{Spintronics Research Network Division, Institute for Open and Transdisciplinary Research Initiatives,Osaka University, Yamadaoka 2-1, Suita, Osaka, 565-0871, Japan}

\date{\today}

%-----------------abstract -------------------
\begin{abstract}
We report the ground-state symmetry of the Ce $4f$ states in the heavy fermion superconductor CeNi$_{2}$Ge$_{2}$, yielding anisotropic $cf$-hybridization between the Ce $4f$ states and conducting electrons. 
By analyzing linear dichroism in soft x-ray absorption and core-level hard x-ray photoemission spectra, the $4f$ symmetry is determined as $\Sigma$-type $\Gamma_{7}$,  promoting predominant hybridization with the conducting electrons originating from the Ge site.
The band structures probed by the soft x-ray angle-resolved photoemission indicates that the Ge $4p$ components contribute to the band renormalization through the anisotropic hybridization effects, suggesting that the control of the electronic structures of Ge orbital gives an impact to achieve the exotic phenomena in CeNi$_{2}$Ge$_{2}$. 
\end{abstract}

\maketitle

\section{INTRODUCTION}

Quantum critical (QC) phenomena such as unconventional superconductivity \cite{CeCu2Si2, SC} with an enormous effective mass enhancement \cite{CeAl3,CeCu2Si2, Cu1, SC, CeSH, RuSH} has been one of the central topics in the strongly correlated 4$f$-based heavy fermion (HF) systems.
Moreover, the intriguing new phenomena such as topological phases and competition or cooperation between magnetism and superconductivity have been reported in this decades~\cite{Cu2, CeRhIn5, CeIn3, Yashima}.
The origin of the QC phenomena for realistic HF systems is proposed by several models based on the spin fluctuation~\cite{Moriya}, Kondo breakdown~\cite{break1,break2,Si}, and valence fluctuation theories~\cite{watanabe1,watanabe2}. 
In addition, it has been pointed out that the anisotropic hybridization effects between the conduction bands and the localized 4$f$ states with the highly anisotropic charge distribution are essential to drive the exotic phenomena~\cite{Hansmann2008, Willers2012, Willers2015}.

CeNi$_{2}$Ge$_{2}$, which is one of Ce-based ternary compounds with the tetragonal ThCr$_{2}$Si$_{2}$ structure, shows non-Fermi liquid behavior with superconducting phase transition below 0.2 K at ambient pressure~\cite{SC}. 
Sommerfeld coefficient is evaluated as 350 mJ/(mol K$^{2}$) \cite{CeSH}, suggesting that the ground states are in the vicinity of QC point. 
The electronic structures have been investigated by angle resolved photoemission (ARPES) in the three-dimensional reciprocal space~\cite{Ehm2001,NakataniPRB2018}, showing the band renormalization at the particular momenta~\cite{NakataniPRB2018}. 
Since anisotropy of the band renormalization was originating from the anisotropic $cf$-hybridization between the local $4f$ orbitals and itinerant valence bands, this motivates us the detailed investigation of the $4f$ symmetry on CeNi$_{2}$Ge$_{2}$ to reveal the origin of the unconventional QC phenomena in the ground states. 

In the tetragonal crystalline electric field (CEF), the local $4f$ states with a total angular momentum $J = 5/2$ for Ce$^{3+}$ ions are expressed by three Kramers doublets equated as follows,
$\ket{\Gamma_{7}^{1}} = \alpha \left|J_{z}=\pm \frac{5}{2}\right> + \sqrt{1-\alpha^{2}}\left|\mp \frac{3}{2}\right>$, 
$\ket{\Gamma_{7}^{2}} = \sqrt{1-\alpha^{2}} \left|\pm \frac{5}{2}\right> - \alpha\left|\mp \frac{3}{2}\right>$, and
$\ket{\Gamma_{6}} = \left|\pm \frac{1}{2}\right>$, 
where the parameter of $\alpha$ ($-1\leqq\alpha\leqq1$) and its $\pm$ sign gives the $c$-axis anisotropy and the in-plane rotational symmetry of the $\Gamma_{7}$ states, respectively. 
To determine the symmetry of the CEF-split $4f$ ground state in the tetragonal Ce compounds, it is powerful to utilize Ce $3d$-$4f$ x-ray absorption spectroscopy (XAS) and Ce $3d$ core-level hard x-ray photoemission (HAXPES).
The dipole-allowed selection rules for the linearly polarized x-ray work in the optical processes of XAS and HAXPES, which helps us to determine the $4f$ symmetry not only for the $c$-axis anisotropy but also for the in-plane rotational symmetry~\cite{CeAgSb2XMCD,ArataniPRB2018,FujiwaraJPSconfProc2020}.

In this paper, we report the symmetry of the Ce $4f$ states in the CEF-split ground-state by the combined spectroscopic technique for CeNi$_{2}$Ge$_{2}$ to reveal the origin of the anisotropic $cf$-hybridization.
The magnetic properties in the Ce $4f$ ground states are discussed by using the x-ray magnetic circular dichroism (XMCD) in XAS spectra in the Ce $M_{4,5}$ edges.
Moreover, the  soft x-ray ARPES probes the detailed analyses of the band structures and Fermi surfaces reflecting the anisotropic hybridization effects.

\section{Experiment}
High quality $R$Ni$_{2}$Ge$_{2}$ ($R$ = Ce, La) single crystals were grown by the Czochralski method. 
The spectroscopic experiments were conducted at the measurement temperature below 20 K that is sufficiently lower than the excited states above 200 K~\cite{Kuwai}.
The linearly and circularly polarized XAS measurements were performed at BL27SU and BL23SU of SPring-8, respectively, where the absorption spectra were obtained in total-electron-yield mode~\cite{BL27SU1, BL27SU2,figure-8,Saitoh}.
The XMCD spectra were measured under magnetic fields up to 10 T along the incident beam direction in 1 Hz helicity-switching mode~\cite{Saitoh}. 
The HAXPES measurements were performed at BL19LXU of SPring-8, where the double diamond phase retarders were installed to switch the linear polarization of the incident x-ray with the photon energy of 7.9 keV~\cite{YabashiPRL2001,Fujiwara2016}.
Since a MBS A1-HE hemispherical photoelectron spectrometer was installed in the horizontal plane with an angle to incident photons of 60$^{\circ}$, the experimental configuration for the horizontally (vertically) polarized light corresponds to the p-polarization (s-polarization), respectively. 
In addition, the detection direction of the photoelectrons with respect to the crystal axes were optimized by the two-axis manipulator with polar ($\theta$) and azimuthal ($\phi$) rotation angles~\cite{Fujiwara2016}. 
The energy resolution was set to about 550 meV for the HAXPES experiments.
Soft x-ray ARPES experiments were performed at BL23SU of SPring-8 \cite{Saitoh} using a Gammadata-Scienta SES-2002 electron analyzer. 
The energy resolution was set to about 70-130 meV for photon energy $(h\nu)$ of about 580-780 eV for ARPES experiments. 
The samples were cleaved \textit{in situ} to expose clean (001) surfaces. 
Moreover, sample quality was checked by the absence of O and C 1$s$ core-level peaks in the photoemission experiments, which are derived from possible impurities or surface oxidization~\cite{NakataniPRB2018}.

\section{RESULTS AND DISCUSSION}

\subsection{Ground state $4f$ symmetry}

\begin{figure}
\begin{center}
\includegraphics[width=8.2cm]{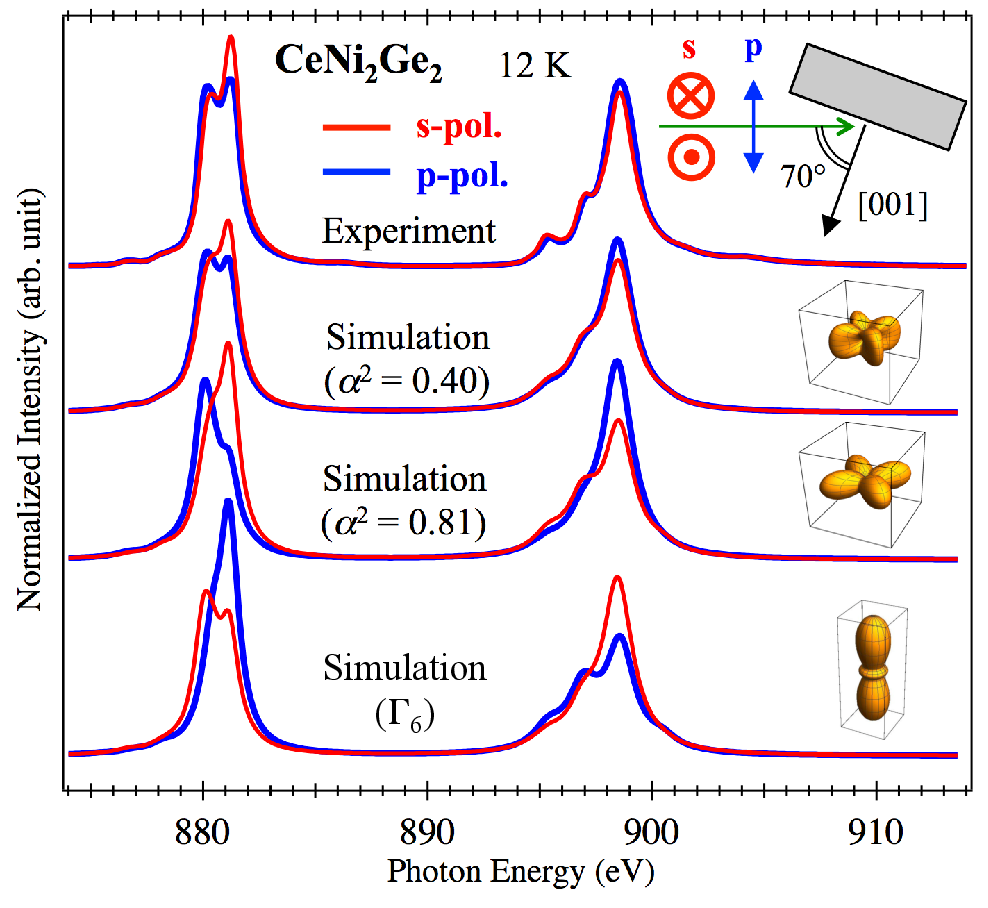}
\end{center}
\begin{center}
\caption{
Linearly polarized XAS spectra of CeNi$_{2}$Ge$_{2}$ at Ce $M_{4,5}$ edges for a photon incidence angel $\theta$ = 70$^{\circ}$ with s- and p-polarization, together with the spectral simulations assuming the ground state in the $\Gamma_{7}$ ($\alpha^{2}$ = 0.4 and 0.81) symmetry and that in the $\Gamma_{6}$ symmetry. 
}
\label{fig:XASLD}
\end{center}
\end{figure}

Figure \ref{fig:XASLD} shows the XAS spectra of CeNi$_{2}$Ge$_{2}$ obtained at Ce $M_{4,5}$ edges with a photon incidence angel $\theta$ = 70$^{\circ}$.
The spectra show the linear polarization dependence for the s(p)-polarized configurations, denoted as s-pol. (p-pol.), respectively, reflecting the anisotropic orbital symmetry of the Ce $4f$ states along the $c$-axis.   
To extract the information of the $c$-axis anisotropy, we have performed ion-model calculations for the XAS spectra including the full multiplet theory implemented by XTLS ver. 9.0 program~\cite{Tanaka1994}.
The atomic parameters were obtained from the Hartree-Fock values for Ce$^{3+}$ ion~\cite{,atomicCalcPar,TholeXASLD}. 
By comparing the spectral simulation, we can clearly exclude the possibilities of the $\Gamma_{6}$ symmetry in the ground states. 
Moreover, we found the anisotropic parameter $\alpha^{2}$ of 0.4 for the $\Gamma_{7}$ states using the ionic model calculations to fit the linearly polarized XAS spectra. 
Note that the simulation with $\alpha^{2} = 0.81$, which was given by the magnetic susceptibility~\cite{AokiXASLD}, cannot explain the experimental results.

%-----------------------%-----------------------%-----------------------%-----------------------%-----------------------

%-----------------------%-----------------------%-----------------------%-----------------------%-----------------------

\begin{figure}
\begin{center}
\includegraphics[width=8cm]{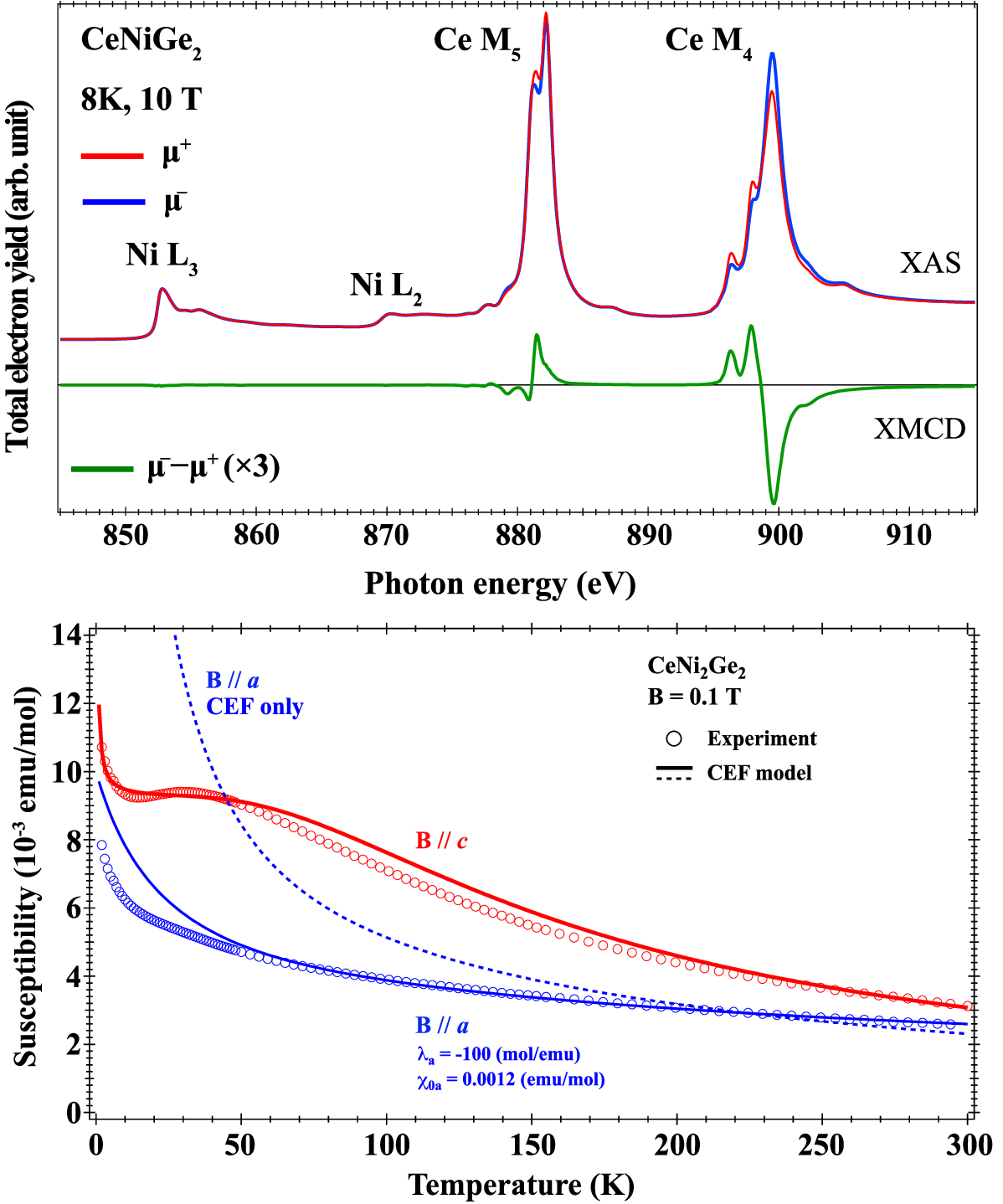}
\end{center}
\begin{center}
\caption{
{
(a) XAS and XMCD spectra at Ni $L_{2, 3}$ and Ce $M_{4, 5}$ edges of CeNi$_{2}$Ge$_{2}$. 
(b) Magnetic susceptibility of CeNi$_{2}$Ge$_{2}$. 
Dotted data indicate the experimental results~\cite{Miyake2017} with the simulation (solid lines) using the parameters in Table \ref{table: Magn}, (see APPENDIX).
}
}
\label{fig:XMCD}
\end{center}
\end{figure}

%-----------------------%-----------------------%-----------------------%-----------------------%-----------------------

\begin{figure}[h]
\begin{center}
\includegraphics[width=7.5cm]{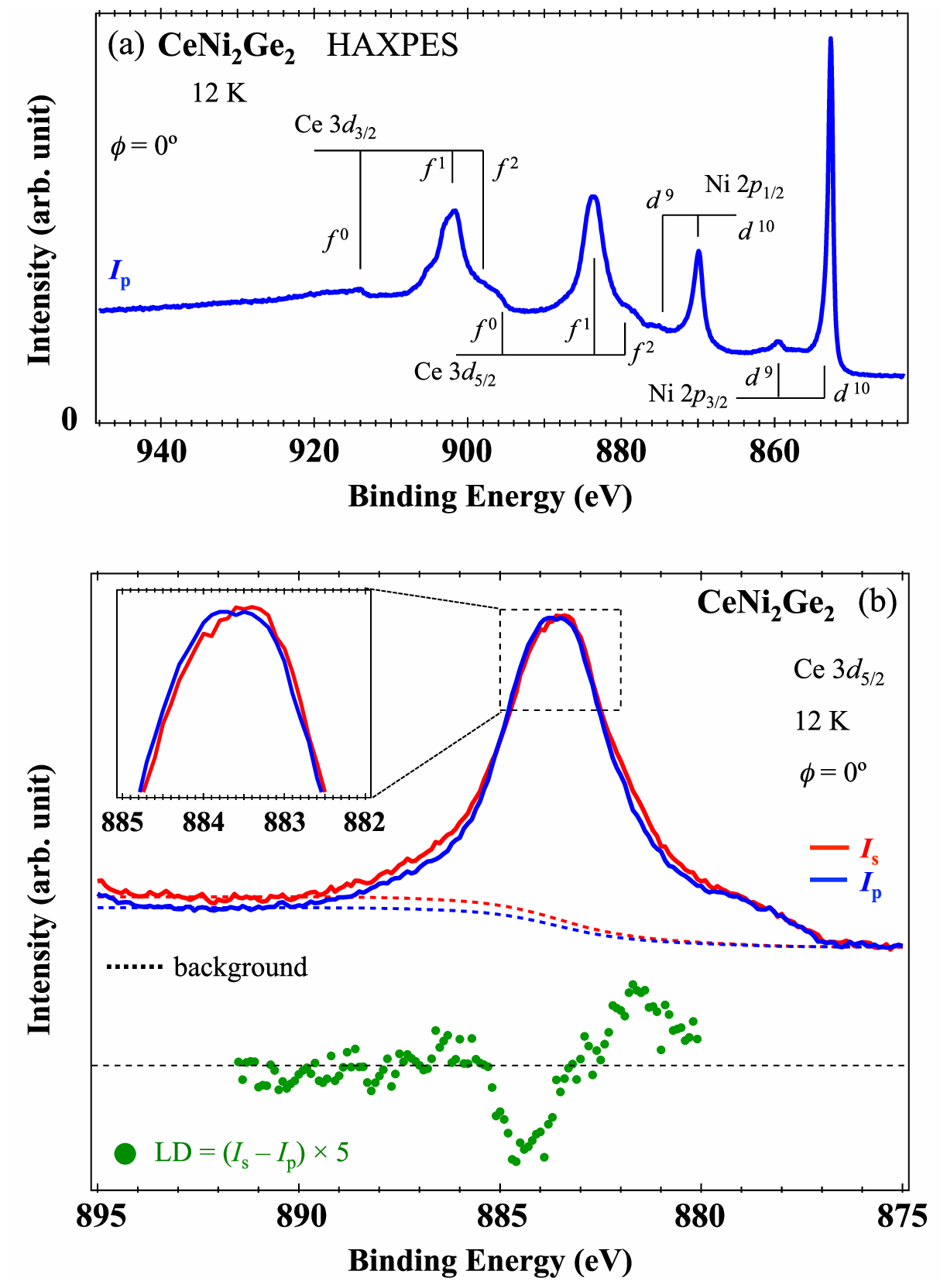}
\end{center}
\begin{center}
\caption{
(a) Ce 3$d$ core-level spectrum together with Ni 2$p$ states. (b) Ce 3$d_{5/2}$ state (solid lines) with the subtracted background (dashed lines).
}
\label{fig:HAXPESBG}
\end{center}
\end{figure}

%-----------------------%-----------------------%-----------------------%-----------------------%-----------------------

\begin{figure*}
\begin{center}
\includegraphics[width=17cm]{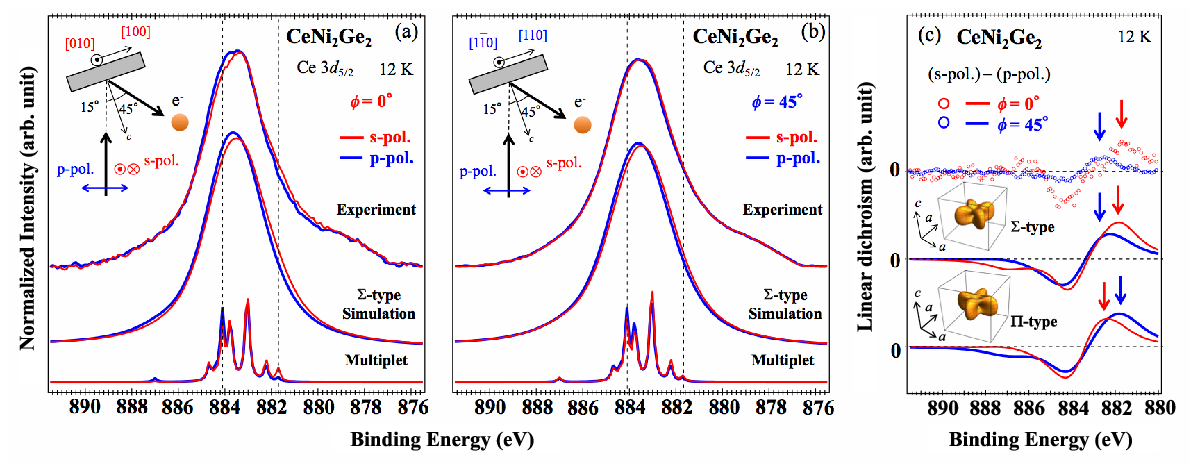}
\end{center}
\begin{center}
\caption{
(a), (b) Ce 3$d_{5/2}$ core level photoemission spectra with s- and p-polarization, together with the spectral simulations and multiplet calculations assuming the ground state symmetry of $\Sigma$-type $\Gamma_{7}$. The photon incidence and photoelectron emitting angels are 15$^{\circ}$, 45$^{\circ}$, respectively. The azimuthal angle ($\phi$) is set to be 0$^{\circ}$ (a) and 45$^{\circ}$ (b) for the crystal $c$-axis. The experimental geometries are illustrated in the insets. 
Note that the spectra obtained in the 0$^{\circ}$ configuration ($\Delta$E$\sim400$ meV) in (a) was recorded slightly better energy resolution than that for the 45$^{\circ}$ configuration ($\Delta$E$\sim550$ meV) in (b). 
The dashed lines in (a) and (b) are guide to the eye to stress the multiplet structures. 
(c) Azimuthal ($\phi$) dependence of the LD spectra obtained from the experiment and the spectral simulations with $\Sigma$- and $\Pi$-types.}
\label{fig:HAXPESLD}
\end{center}
\end{figure*}
%-----------------------%-----------------------%-----------------------%-----------------------%-----------------------

Validity of the anisotropic parameter obtained from the spectroscopy can be checked by the local magnetic moment.
However, it is not simple for CeNi$_{2}$Ge$_{2}$, which is constituted by two magnetic elements of Ce and Ni. 
The element specific investigation using XMCD in both Ce and Ni edges is thus important to discuss the magnetic contribution derived from the Ce 4$f$ and Ni 3$d$ states~\cite{CTChenXMCD,ElementXMCD,CeAgSb2XMCD,CeFe2XMCD}. 
Figure \ref{fig:XMCD}(a) shows the XAS spectra recorded in parallel ($\mu^{+}$) and antiparallel ($\mu^{-}$) configurations between the light helicity and the direction of the magnetic field.
The circular polarization dependence is clearly observed in the Ce $M_{4,5}$ edges as highlighted in the XMCD spectra defined as $\mu^{-}-\mu^{+}$.
Meanwhile, the XMCD signals are absent in the Ni $L_{2,3}$ edges, indicating that the Ni $3d$ states does not contribute to the magnetic properties in CeNi$_{2}$Ge$_{2}$.

Quantitative information on the total atomic magnetic moment of the Ce $4f$ states can be obtained by applying the corrected XMCD sum rules~\cite{TholeXMCD, CarraXMCD, TeramuraXMCD},
\begin{align}
\langle{L_{z}}\rangle &= \frac{2q(14-n_{f})}{rP_{c}}, \\
\frac{\langle{L_{z}}\rangle}{\langle{S_{z}}\rangle} &= \frac{4C}{5p/q-3}\left(1+3\frac{\langle{T_{z}}\rangle}{\langle{S_{z}}\rangle}\right),
\end{align}
where $p$ ($q$) is the integral of the XMCD signal over the $M_{5}$ ($M_{4, 5}$) edges, $r$ is the integral of the polarization-summed XAS intensity over the $M_{4, 5}$ edges, $P_{c}$ denotes degree of circular polarization of the incident x-rays, and $C$ is the correction factor for the mixing of the multiplet structure between the 3$d_{5/2}$ and 3$d_{3/2}$ levels caused by 3$d$-4$f$ electrostatic interactions. 
$n_{f}$ is the occupation number of 4$f$ electrons, and $\langle{T_{z}}\rangle$ is the expectation value of the magnetic dipole operator. 
Assuming $P_{c}$ = 0.97~\cite{Saitoh,CeAgSb2XMCD}, $n_{f}$ = 1, and an atomic $\langle{T_{z}}\rangle/\langle{S_{z}}\rangle$ ratio of 8/5 with $C$ = 1.6~\cite{TeramuraXMCD}, 
we obtained the total magnetic moment of Ce$^{3+}$ ion as 0.092 $\mu_{{\rm B}}$. 
The bulk magnetization measurements suggest the magnetization of CeNi$_{2}$Ge$_{2}$ is around 0.1 $\mu_{{\rm B}}$ at 10 T and 4.2 K~\cite{Fukuhara1996}. 
Thus, the magnetization of CeNi$_{2}$Ge$_{2}$ is quantitatively explained with the Ce$^{3+}$ ion.

Figure \ref{fig:XMCD}(b) shows the magnetic susceptibility obtained under an external magnetic field of B = 0.1 T parallel to the $a$ and $c$-axes~\cite{Miyake2017}, compared with the simulation based on the CEF model for the Ce$^{3+}$ ion using the CEF parameters as summarized in APPENDIX. 
The $c$-axis susceptibility is well explained by the simulation based on our CEF model, while the $a$-axis susceptibility shows the deviation from the pure CEF calculation. 
Then, the simulated $a$-axis susceptibility is corrected by the additional molecular field $\lambda_{a}$ as follows, 
\begin{align}
\chi^{-1} = (\chi_{{\rm CEF}} + \chi_{0})^{-1} - \lambda_{a},
\label{eq:magIS}
\end{align}
where $\lambda_{a}$ of $-100$ mol/emu, which is comparable to CeCu$_{2}$Ge$_{2}$ with $\lambda_{a}$ of $-50$ mol/emu~\cite{ArataniPRB2018}, and $\chi_{0}$ of $1.2 \times 10^{-3}$ emu/mol are set to reproduce the experimental data along the $a$-axis as shown in the solid line in Fig. \ref{fig:XMCD}(b). 
This is possibly due to a short-range magnetic fluctuation originating from the antiferromagnetic correlation~\cite{Frost2000,KadowakiCEF}. 
Note that the magnetic correlation should be considered only for the in-plane direction since the additional tuning of the molecular field is needed only for the $a$-axis.

Moreover, in-plane symmetry of the $4f$ orbital is probed by using linear polarization dependence of the Ce $3d$ HAXPES spectra~\cite{ArataniPRB2018}.
Figure \ref{fig:HAXPESBG}(a) shows the overview of the core-level HAXPES spectrum for the Ce $3d$ and Ni $2p$ states in CeNi$_{2}$Ge$_{2}$. 
The Ce $3d$ spectrum show the atomic-like structures due to the $3d^{9}4f^{1}$ and  $3d^{9}4f^{2}$ final states, and the tiny hump structure due to the $3d^{9}4f^{0}$ final states, reflecting the sizable $cf$-hybridization effects~\cite{yano2008,Sundermann2016}. 
On the other hand, the sharp peaks due to the $3d^{10}$ states are observed in the Ni $2p$ spectrum with the satellite structures due to the $3d^9$ states as seen in Ni metal~\cite{Huefner1975,vdLaan1992,sueda_86_2017}. 
The peak width of the Ce $3d$ spectrum is wider than that of the Ni $2p$ spectrum originates from the multiplet structures due to the anisotropic Coulomb and exchange interactions between the Ce $3d$ core holes and the $4f$ electrons, giving the linear polarization dependence in the spectra thanks to the dipole selection rules. 
In Fig. \ref{fig:HAXPESBG}(b), the Ce $3d_{5/2}$ spectra obtained s- and p-polarized configuration, denoted as $I_\textrm{s}$ and $I_\textrm{p}$, clearly show the linear polarization dependence as shown in the inset. 
Then, the difference of normalized intensity $I_\textrm{s}-I_\textrm{p}$ after subtracting the background is defined as the linear dichroism (LD), reflecting the anisotropic charge distribution of the 4$f$ states~\cite{ArataniPRB2018}.

To discuss the in-plane symmetry, the Ce $3d_{5/2}$ spectra are recorded at two different arrangements by rotating the azimuthal angle $\phi$ between 0$^{\circ}$ and 45$^{\circ}$ to change the detection directions of the photoelectrons as shown in Fig.~\ref{fig:HAXPESLD}(a) and (b), respectively.
The linear polarization dependence at around 884.5 eV in the $\phi= 0^{\circ}$ configuration is stronger than that in the $\phi= 45^{\circ}$ configuration, which is due to the intensity difference of the multiplet structures as simulated by using the ionic calculation~\cite{Tanaka1994,atomicCalcPar}. 
This is highlighted in the azimuthal angle dependence of the LD spectra obtained in the $\phi=0^{\circ}$ and 45$^{\circ}$ configurations as shown in Fig.~\ref{fig:HAXPESLD}(c).
Moreover, the LD spectrum for the $\phi= 0^{\circ}$ configuration also shows the positive peak at around 882 eV, which is located to the lower binding energy side than that for the $\phi= 45^{\circ}$ configuration.
The azimuthal angle dependence of LD signals gives a good fingerprint of the in-plane symmetry of the $\Gamma_{7}$ ground states, which is characterized by the $\pm$ sign of the anisotropy parameter $\alpha$ contributing to the linear combination between $\left|5/2\right>$ and $\left|3/2\right>$ components. 
The ionic calculations for the Ce $3d$ photoemission explain the experimental LD signals when the $\Sigma$-type $\Gamma_{7}$ with $\alpha = -0.4$. 
$\Pi$-type $\Gamma_{7}$ ($\alpha>0$) is clearly ruled out since the positions of the positive peaks around 882 eV are reversed in the LD spectra for $\phi= 0^{\circ}$ and $45^{\circ}$ configurations.

\begin{figure}
\begin{center}
\includegraphics[width=8.5cm]{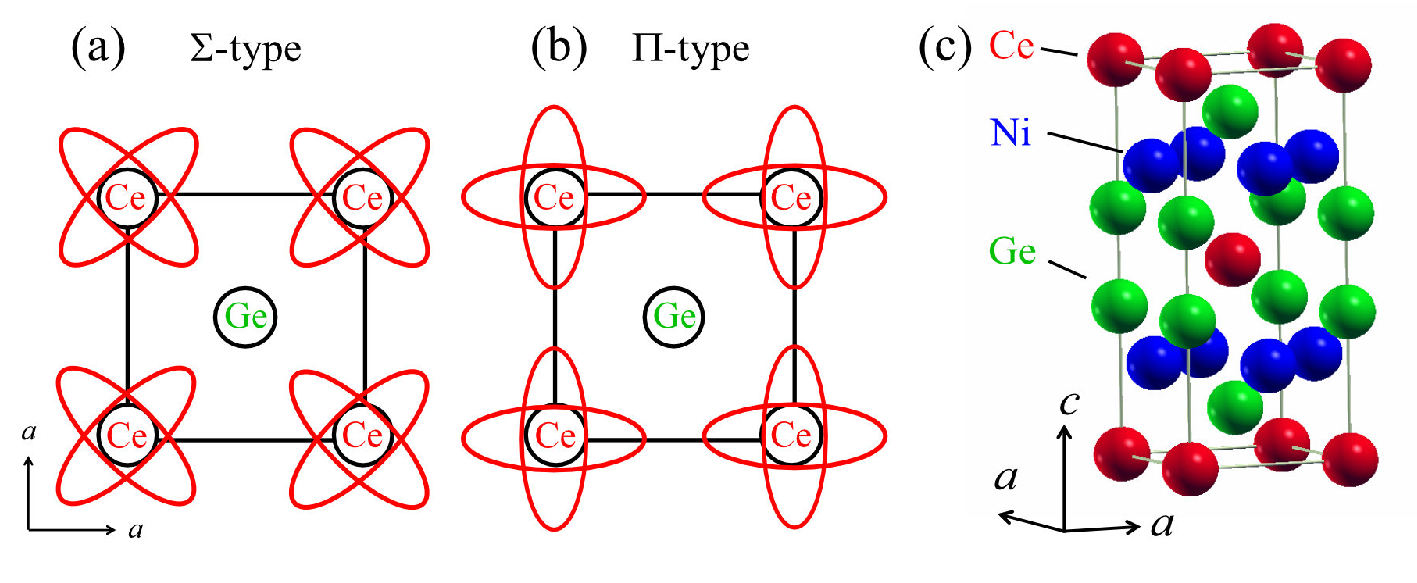}
\end{center}
\begin{center}
\caption{
The schematic view of charge distribution for $\Sigma$- and $\Pi$-type $\Gamma_7$ orbital of the Ce $4f$ states on CeNi$_{2}$Ge$_{2}$ in the $ab$-plane of the tetragonal lattice in (a) and (b), respectively, together with the crystal structure of CeNi$_{2}$Ge$_{2}$ in (c).
}
\label{fig:SigmaPi}
\end{center}
\end{figure}

By utilizing combined core-level spectroscopies we have successfully determined the ground state $4f$ symmetry of CeNi$_{2}$Ge$_{2}$ as
\begin{eqnarray*}
 \ket{\Sigma\mathchar`-{\rm type}~\Gamma_7}=-\sqrt{0.4}\ket{\pm \frac{5}{2}}+\sqrt{0.6}\ket{\mp \frac{3}{2}}.
\label{eq:HAXPESGS}
\end{eqnarray*}
The 4$f$ states with $\Sigma$-type $\Gamma_7$ symmetry have charge distribution to the corner of the unit cell as illustrated in Fig.~\ref{fig:SigmaPi}, which is pointed to the Ge site.
Therefore, it is promoted that the Ce $4f$ states hybridized with conduction electrons originating from the nearest neighbor Ge sites, while the relatively weak $\pi$-bonding-like hybridization is realized between the Ce and Ni sites.

%-----------------------%-----------------------%-----------------------%-----------------------%-----------------------

%-----------------------%-----------------------%-----------------------%-----------------------%-----------------------

\subsection{Valence-band electronic structures}

\begin{figure}
\begin{center}
\includegraphics[width=7.5cm]{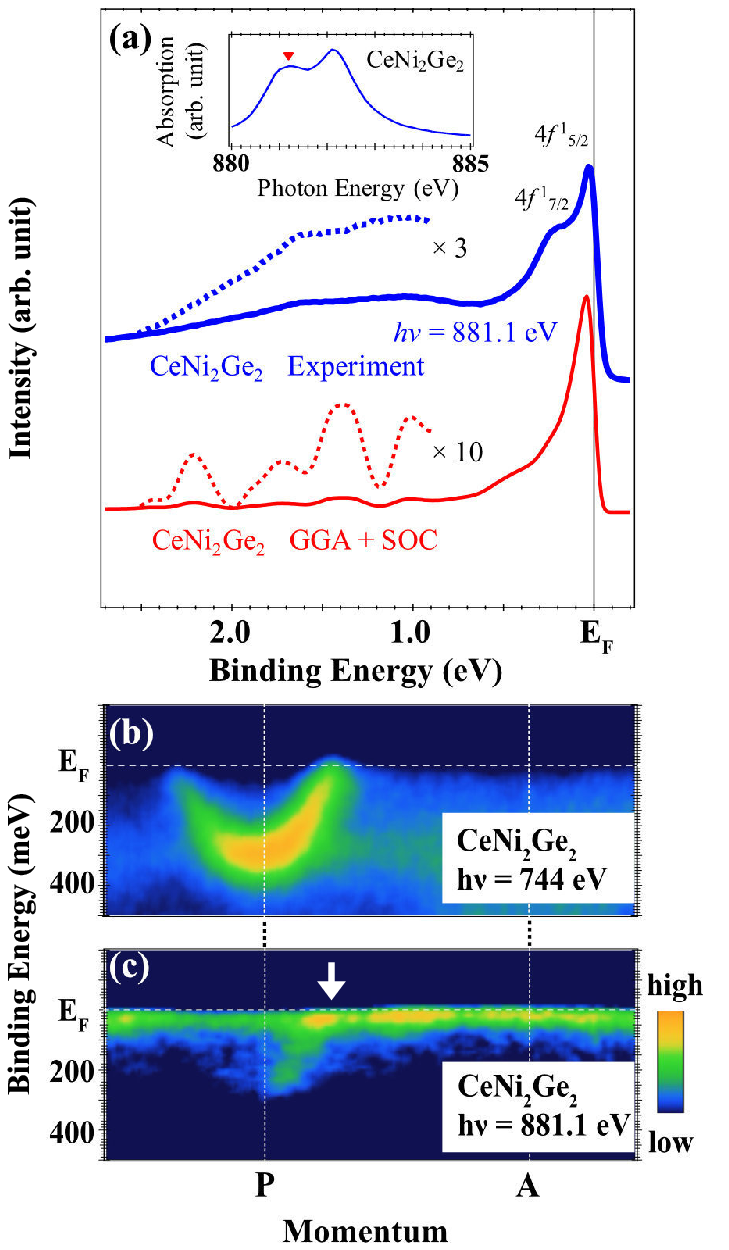}
\end{center}
\begin{center}
\caption{
(Color online)  (a) Ce 3$d$-4$f$ resonance photoemission spectra of CeNi$_{2}$Ge$_{2}$. 
The inset shows the X-ray absorption spectrum of CeNi$_{2}$Ge$_{2}$ with the marker indicating the Ce 3$d$-4$f$ resonance condition. 
The $4f$ PDOS convoluted by Fermi-Dirac function and the energy resolution is plotted in the bottom.
(b), (c) ARPES intensity plots of CeNi$_{2}$Ge$_{2}$ obtained by the photon energies of 744 and 881.1 eV, respectively.
}
\label{fig:ResPES_ARPES}
\end{center}
\end{figure}

To discuss further details of the anisotropic hybridization effects, it is important to investigate the valence band electronic structures.
Figure~\ref{fig:ResPES_ARPES}(a) displays the Ce $3d$-$4f$ resonant photoemission (RPES) spectra obtained at $h\nu = 881.1$ eV located at the Ce $M_{5}$ absorption peak as indicated in the inset to enhance the Ce $4f$ states contributions~\cite{CeAgSb2XMCD,Sekiyama2000}. 
The spectrum shows the sharp peak at around 0.1 eV originates from the Ce $4f^{1}_{5/2}$ contributions with contributing to the Fermi level ($E_\textrm{F}$), representing the tail of the Kondo-resonance peak in the Ce $4f$ states, and the  Ce $4f^{1}_{7/2}$ final states are observed in the shoulder structure at $\sim$0.3 eV as the spin-orbit partner. 
The spectra are compared with the simulation based on the itinerant model of the Ce $4f$ electrons, which is obtained by the partial density of states (PDOS) of the Ce $4f$ components convoluted by Fermi-Dirac function and the energy resolution for the RPES experiments of 60 meV. 
The band structure calculation was implemented by WIEN2k package~\cite{WIEN2kRPES,WIEN2k_setup}. 
The generalized gradient approximation (GGA) using the Perdew-Burke-Ernzerhof scheme has been used for the exchange correlation potential~\cite{PBE1RPES, PBE2RPES}, and the spin-orbit coupling (SOC) was included for the Ce 4$f$ states in the scalar-relativistic scheme. 
In the simulated spectra, the Ce $4f^{1}_{7/2}$ contribution observed in the experiment is absent, indicating that the Ce $4f$ states have a degree of localized character, which is consistent with the results of the core-level spectroscopies as discussed in the former section.

%-----------------------%-----------------------%-----------------------%-----------------------%-----------------------
%-----------------------%-----------------------%-----------------------%-----------------------%-----------------------

%-----------------------%-----------------------%-----------------------%-----------------------%-----------------------

\begin{figure*}
\begin{center}
\includegraphics[width=12cm]{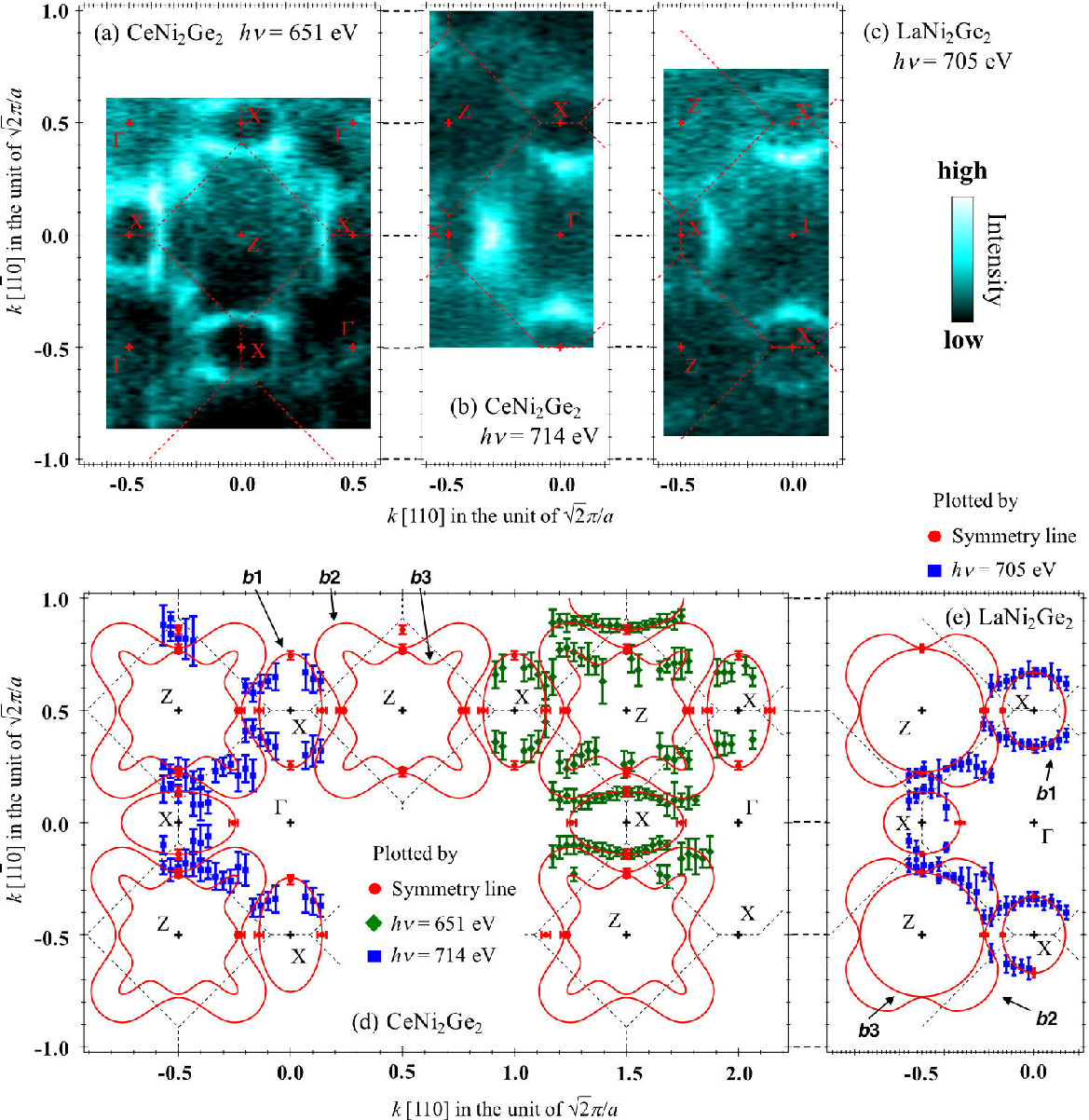}
\end{center}
\begin{center}
\caption {
FS slices in the $k_{x}$-$k_y$ planes of CeNi$_{2}$Ge$_{2}$ and LaNi$_{2}$Ge$_{2}$. 
(a)-(c) Momentum distributions of ARPES intensity in the Z-centered (a) and $\Gamma$-centered (b), (c) plane of CeNi$_{2}$Ge$_{2}$ (a), (b) and LaNi$_{2}$Ge$_{2}$ (c), which are obtained by integrating the photoelectron intensity from the Fermi level and $-0.2$ eV in the unoccupied side. 
The dashed lines represent the BZ boundaries. 
(d), (e) $k_{{\rm F}}$ plots for CeNi$_{2}$Ge$_{2}$ (d) and LaNi$_{2}$Ge$_{2}$ (e). 
The dots with error bars represent the $k_{{\rm F}}$'s estimated from each specific angle slice obtained by the 651 and 714 eV data together with the high statistics measurements along the high symmetry lines. 
The solid lines are guides to the eye of Fermi surfaces following the experimentally evaluated $k_{{\rm F}}$'s. The dashed lines represent the BZ boundaries.
}
\label{fig_FS}
\end{center}
\end{figure*}

%-----------------------%-----------------------%-----------------------%-----------------------%-----------------------

Figures~\ref{fig:ResPES_ARPES}(b) and (c) show the ARPES spectra obtained at off-resonance ($h\nu=744$ eV) and on-resonance ($h\nu=881.1$ eV) conditions along the P-A line in the three-dimensional reciprocal lattice, respectively~\cite{NakataniPRB2018}.
The parabolic band structures around P point are clearly observed in the off-resonance ARPES spectra in Fig.~\ref{fig:ResPES_ARPES}(b).
These bands are also observed in the on-resonance ARPES spectra as shown in Fig.~\ref{fig:ResPES_ARPES}(c), originating from the possible Ce $5d$ components, which are also enhanced at the Ce $3d$-$4f$ resonance~\cite{MYKimura2015}.
In addition, on-resonance ARPES spectra show the weakly dispersed band besides $E_\textrm{F}$ mainly due to the Ce $4f$ states. 
The intensity increases by crossing the parabolic band indicated by the downward arrow in Fig.~\ref{fig:ResPES_ARPES}(c), suggesting the band renormalization due to the $cf$-hybridization, although the energy resolution is not sufficient for the further detailed discussion.
From our previous report on CeNi$_2$Ge$_2$~\cite{NakataniPRB2018}, the strong band renormalization was clearly observed in the band structures in the $\Gamma$-X line, which was measured at the different photon energy with the off-resonance condition, i.e., different $k_z$ for the P-A line. 
Thus, we should focus on the detailed electronic structures originating from the conduction electrons in the Z-$\Gamma$-X plane in the reciprocal space to elucidate the anisotropic $cf$-hybridization effects.

%-----------------------%-----------------------%-----------------------%-----------------------%-----------------------

\begin{figure}
\begin{center}
\includegraphics[width=8.5cm]{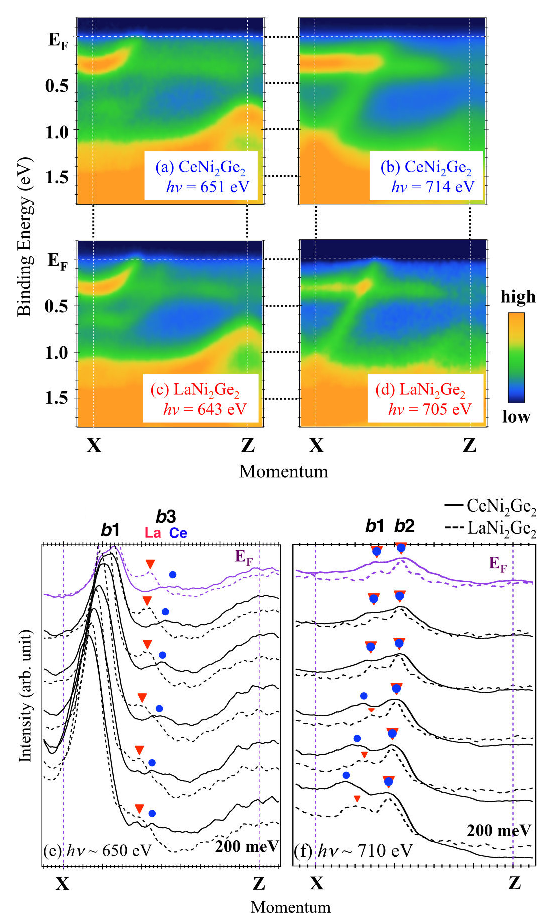}
\end{center}
\begin{center}
\caption{
(a), (b) ARPES intensity plot of CeNi$_{2}$Ge$_{2}$ along the Z-X direction recorded with $h\nu$ = 651 (a) and 714 (b) eV. (c), (d) Same as (a), (b) of LaNi$_{2}$Ge$_{2}$ recorded with $h\nu$ = 643 (c) and 705 (d) eV. 
Note that the ARPES data obtained at the different photon energies are measured at different polar angles to measure the same symmetry lines in the Brillouin zone of the body-centered tetragonal lattice. 
(e), (f) MDCs of CeNi$_{2}$Ge$_{2}$ (solid lines) and LaNi$_{2}$Ge$_{2}$ (dashed lines) with $h\nu$ $\sim$ 650 (e) and 710 (f) eV, which is extracted from (a)-(d).
}
\label{fig_band_ZX}
\end{center}
\end{figure}

%-----------------------%-----------------------%-----------------------%-----------------------%-----------------------

\begin{figure}
\begin{center}
\includegraphics[width=8.5cm]{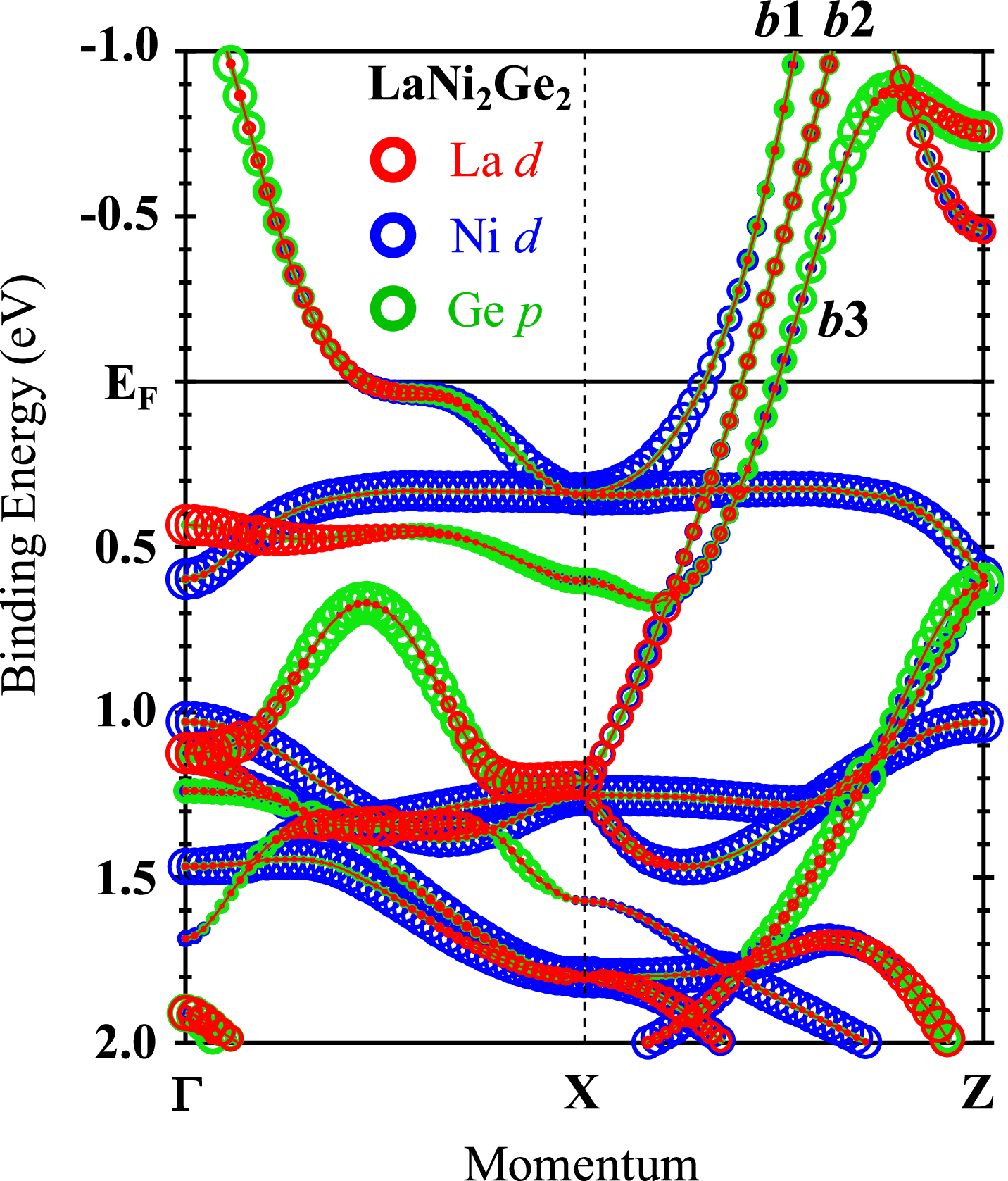}
\end{center}
\begin{center}
 \caption{
Calculated band structures of LaNi$_{2}$Ge$_{2}$. The size of the dots indicates degree of the orbital components of La $5d$,  Ni $3d$, and Ge $4p$ states. 
}
\label{fig:LaNi2Ge2_DFT}
\end{center}
\end{figure}

%-----------------------%-----------------------%-----------------------%-----------------------%-----------------------

Intensity plots of the ARPES spectra at $E_\textrm{F}$ in the Z-$\Gamma$-X reciprocal plane, reflecting the slice of Fermi surfaces (FSs) of CeNi$_2$Ge$_2$, are obtained at the two different photon energies of 651 eV and 714 eV as shown in Figs.~\ref{fig_FS}(a) and (b), respectively. 
The difference of the intensity distribution for both photon energies is mainly due to the matrix element effects, suggesting the difference of symmetry of the bands crossing $E_\textrm{F}$~\cite{Moser2017}. 
For both photon energies, the ellipsoidal FSs are clearly observed around the X points. 
Especially, the FSs observed in 714 eV are elongate to the $\Gamma$ point. 
Comparing to the mapping on LaNi$_2$Ge$_2$, which is a good reference without $4f$ electrons to probe the band structures due to the conduction electrons, Fermi momentum ($k_\textrm{F}$) is closer to the $\Gamma$ point in CeNi$_2$Ge$_2$ as shown in Fig.~\ref{fig_FS}(c), suggesting the anisotropic $cf$-hybridization effects. 
This is highlighted by plotting $k_\textrm{F}$ positions estimated from the momentum distribution curves (MDCs) as shown in Figs.~\ref{fig_FS}(d) and (e) for CeNi$_2$Ge$_2$ and LaNi$_2$Ge$_2$, respectively.
The FSs around X points are clearly elongated to the $\Gamma$ points in the CeNi$_2$Ge$_2$, indicating that the renormalized band is only observed in the band structures along the $\Gamma$-X-$\Gamma$ lines in CeNi$_2$Ge$_2$, which is consistent with the previous report~\cite{NakataniPRB2018}.

%-----------------------%-----------------------%-----------------------%-----------------------%-----------------------

\begin{figure*}
\begin{center}
\includegraphics[width=18cm]{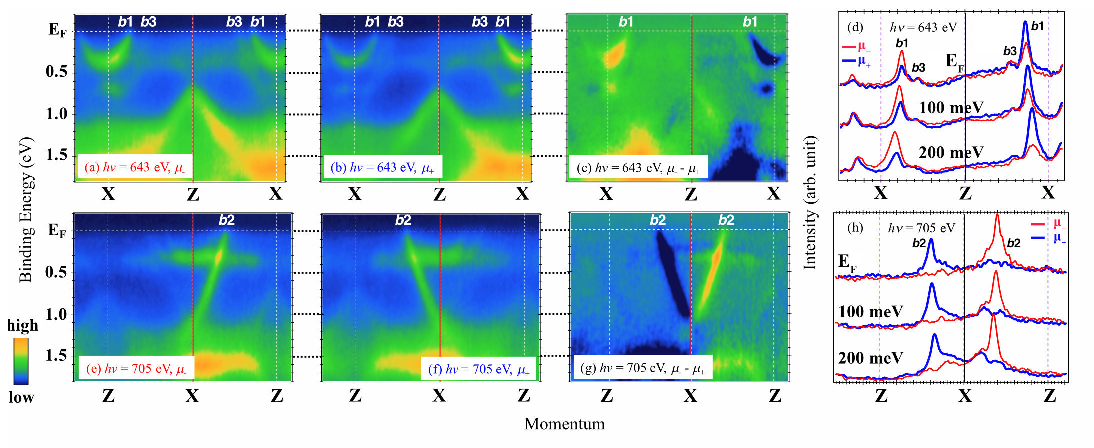}
\caption{
Circular polarized ARPES spectra of LaNi$_{2}$Ge$_{2}$ recorded with $h\nu$ = 643 and 705 eV in (a)-(d) and (e)-(h), respectively. 
ARPES intensity plots are obtained with right ($\mu_{-}$) and left ($\mu_{+}$) circular polarized photons along the Z-X direction in (a), (e) and (b), (f). 
(c), (g) Intensity difference ($\mu_{-}$-$\mu_{+}$) of circularly polarized ARPES spectra. (d), (h) MDCs with right and left circular polarized photons.
}
\label{Fig: CD}
\end{center}
\end{figure*}

%-----------------------%-----------------------%-----------------------%-----------------------%-----------------------

Figures~\ref{fig_band_ZX} show the high symmetry cuts along the Z-X direction recorded at $hv = $651 (643) eV and 714 (705) eV for CeNi$_2$Ge$_2$ (LaNi$_2$Ge$_2$), respectively. 
The ARPES data measured at around 650 eV and 710 eV are obtained in the same symmetry lines, but the different bands are pronounced due to the matrix element effect, suggesting the difference of the symmetry and character of the band structures. 
The three bands, labeled as $b1$, $b2$, and $b3$, are observed in the spectra, and the $b1$ and $b3$ ($b1$ and $b2$) bands are clearly observed at the $h\nu\sim$ 650 eV (710 eV) as shown in Figs.~\ref{fig_band_ZX}(a) and (c) (Figs.~\ref{fig_band_ZX}(b) and (d)), respectively.
For both photon energies, the ARPES images of CeNi$_{2}$Ge$_{2}$ is similar to those of LaNi$_{2}$Ge$_{2}$, but the detailed comparison of the MDCs show the peaks of band $b3$ on CeNi$_{2}$Ge$_{2}$ is closer to the Z point as shown in Fig.~\ref{fig_band_ZX}(e), indicating that the hole-like FS of CeNi$_{2}$Ge$_{2}$ [see also Fig.~\ref{fig_FS}(d)] is smaller than that of LaNi$_{2}$Ge$_{2}$ due to the $cf$-hybridization.
On the other hand, the peak positions of MDCs of bands $b1$ and $b2$ in Fig.~\ref{fig_band_ZX}(f) are comparable for both compounds, indicating that the $cf$-hybridization effects are not clearly observed in bands $b1$ and $b2$ within the energy and  momentum resolutions in the experiments. 
These band structures are qualitatively explained by the band structure calculation for LaNi$_{2}$Ge$_{2}$ as shown in Fig.~\ref{fig:LaNi2Ge2_DFT}.
The weight of the La, Ni and Ge contributions are different for each bands proposed in the theory, supporting that three bands have different character and symmetry.

To focus on the character of the band structures for the conducting electrons, the circular polarization dependence of the ARPES spectra on LaNi$_{2}$Ge$_{2}$ are examined along the X-Z-X line on LaNi$_{2}$Ge$_{2}$ as shown in Figs.~\ref{Fig: CD}. 
The band mapping obtained at 643 eV, the intensity modulation for the left ($\mu_+$) and right ($\mu_-$) circularly polarized light is observed in the inner parabolic band around the X point originating from band $b1$, while the outer band derived from band $b3$ does not show the strong polarization dependence. 
On the other hand, the rapid dispersion of band $b2$ show the strong circular polarization dependence for the X point.
These differences are visualized in the circular dichroism of the band mapping defined as the difference between $\mu_+$ and $\mu_-$ as shown in Figs.~\ref{Fig: CD}(c) and (g). 
The strong contrast is observed in band $b2$ as shown in Fig.~\ref{Fig: CD}(g), and the pronounced difference is clearly observed in the intensity of the MDC peaks for band $b2$, which is much stronger than that for band $b1$.
This indicates that all three bands have different orbital character and symmetry, which induces the anisotropic hybridization effects between the Ce $4f$ states and the conducting band~\cite{VyalikhCDAD,NakataniCDAD}.

From our previous work~\cite{NakataniPRB2018}, band $b1$ along $\Gamma$-X direction contributes to hybridization effects with strong band renormalization yielding the heavy spot in the FSs.
Moreover, the $k_\textrm{F}$ of band $b3$ in CeNi$_{2}$Ge$_{2}$ changes for those in LaNi$_{2}$Ge$_{2}$ along the Z-X line as shown in Fig.~\ref{fig_band_ZX}(e).
Comparing to the band structure calculation on LaNi$_{2}$Ge$_{2}$ in Fig.~\ref{fig:LaNi2Ge2_DFT}, the Ge 4$p$ orbitals contribute to band $ b1$ along $\Gamma$-X direction and band $b3$ near $E_\textrm{F}$, which change its electronic structures due to $cf$-hybridization.
On the other hand, the Ge 4$p$ orbitals do not contribute to band $b1$ along the Z-X direction and band $b2$, which do not change its electronic structures between CeNi$_{2}$Ge$_{2}$ and LaNi$_{2}$Ge$_{2}$ as shown in Fig.~\ref{fig_band_ZX}(f).
These facts suggest that the Ge 4$p$ components are important for controlling the electronic structures in CeNi$_{2}$Ge$_{2}$.
We remind that the ground state symmetry of the Ce 4$f$ orbital is the $\Sigma$-type with the charge distribution elongated to the corner of the unit cell as shown in Fig.~\ref{fig:SigmaPi}(a), indicating that the predominant hybridization between the Ce $4f$ states and Ge orbital is preferable.
These results correspond well to the fact that the Ge 4$p$ components are important to control the electronic structures and physical properties, giving the impact to the in-plane magnetic correlations as shown in Fig. 2(b), through the valence electrons derived from Ge, inducing the Ruderman-Kittel-Kasuya-Yosida interaction in CeNi$_{2}$Ge$_{2}$.

\section{CONCLUSION}

By using the core-level spectroscopies we have determined the ground state symmetry of the Ce $4f$ orbital in CeNi$_{2}$Ge$_{2}$ as $\ket{\Sigma\mathchar`-{\rm type}~\Gamma_7}=-\sqrt{0.4}\ket{\pm \frac{5}{2}}+\sqrt{0.6}\ket{\mp \frac{3}{2}}$. 
The magnetic properties are well explained by our crystal field model for the Ce$^{3+}$ ion in the tetragonal symmetry.
Since the in-plane symmetry is $\Sigma$-type, the Ce $4f$ electrons are preferable to hybridize with the Ge $4p$ contributions, which is consistent with the detailed investigation of the band structures for CeNi$_{2}$Ge$_{2}$ and LaNi$_{2}$Ge$_{2}$.
This supports the strategy to control the physical properties of CeNi$_{2}$Ge$_{2}$ by substituting the Ge sites to achieve the new exotic phases around the quantum critical points.

\section*{ACKNOWLEDGEMENTS}

We acknowledge the help and support of Y. Takeda of JAEA during the beam time at SPring-8 BL23SU. 
We also thank S. Fujioka, T. Mori, T. Kadono, and T. Muro for supporting the measurements. 
The XMCD and ARPES measurements were performed under the approval of BL23SU at SPring-8 (Proposal Nos. 2013B3882 and 2014B3882). 
The linearly polarized XAS and core-level photoemission measurements were performed under the approval of JASRI (Proposal Nos. 2014A1023, 2014B1299, 2014B1305, 2015A1533).
This work was supported by a Grant-in-Aid for Scientific Research (JP16H04014, JP18K03512, JP18K03537, JP20K20900, JP20H05271, and JP22K03527), a Grant-in-Aid for Innovative Areas (JP20102003, JP16H01074, and JP18H04317), and Grant-in-Aid for Transformative Research Areas (JP23H04867) from MEXT and JSPS, Japan. 
Y. N. was supported by the Program for Leading Graduate Schools “Interactive Materials Science Cadet Program” and JSPS Research Fellowships for Young Scientists.

\section*{APPENDIX: CRYSTAL FIELD PARAMETERS}

\begin{figure}
\begin{center}
\includegraphics[width=8.5cm]{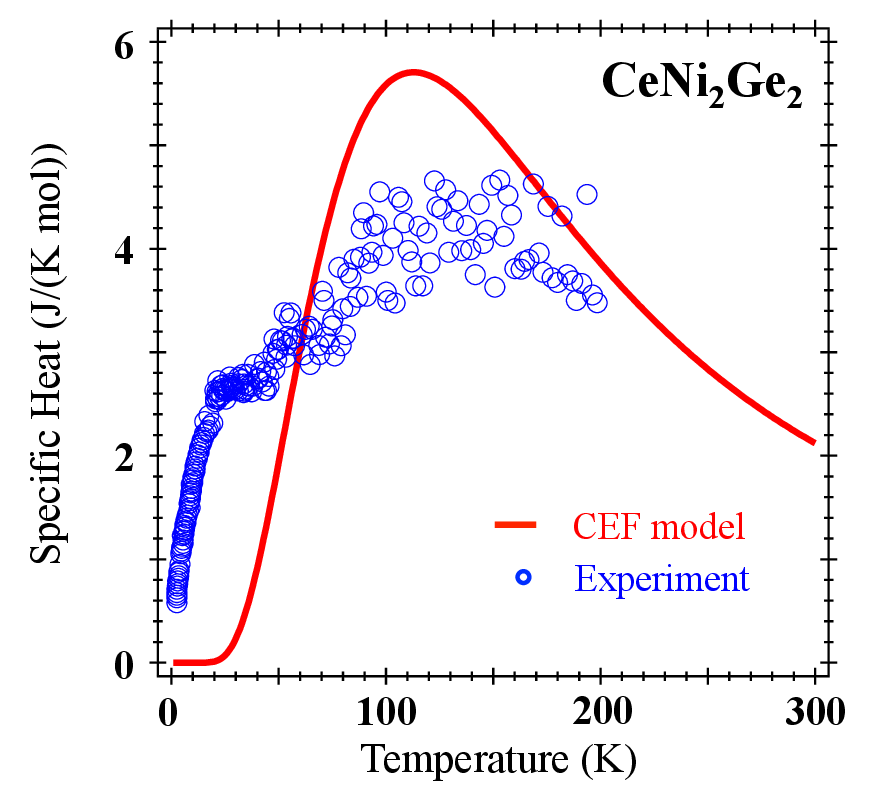}
\end{center}
\begin{center}
\caption{
Specific heat of CeNi$_{2}$Ge$_{2}$. Solid line indicates the simulation based on the CEF model. Dots denote the experimental results~\cite{Kuwai}.
}
\label{fig:SpeficifHeat}
\end{center}
\end{figure}

The crystal field Hamiltonian in the tetragonal symmetry for the Ce$^{3+}$ ion with total angular momentum $J$ = 5/2 states are given by
\begin{align}
H_\textrm{CEF} = B_{2}^{0}O_{2}^{0}+B_{4}^{0}O_{4}^{0}+B_{4}^{4}O_{4}^{4}, 
\end{align}
where $B_{2}^{0}$, $B_{4}^{0}$, and $B_{4}^{4}$ are crystal field parameters for operators $O_{2}^{0}$, $O_{4}^{0}$, and $O_{4}^{4}$ in Stevens formalism~\cite{Stevens1952}. 
To analyze the magnetic susceptibility, we use the crystal field parameters as summarized in Table~\ref{table: Magn} obtained by the anisotropy parameter $\alpha^2 = 0.4$ in the $4f$ symmetry and the energy level of excited states estimated by XAS and specific heat, respectively. 
Figure \ref{fig:SpeficifHeat} shows temperature dependence of the specific heat, which was reported in Ref.~\onlinecite{Kuwai}. 
The broad peak structure around 100-150 K is qualitatively explained by the simulation based on the CEF model, while the structure at $\sim$ 30 K is possibly due to the hybridization effects, which are not directly taken into account in our simulation~\cite{Kuwai}.
By using the simulation based on the crystal field model, we obtained the excited state of 230 K and 390 K, which were reasonably explained by the inelastic neutron scattering~\cite{Frost2000}, as summarized in Table~\ref{table: Magn}. 

%-----------------------%-----------------------%-----------------------%-----------------------%-----------------------
\begin{table}[H]
\begin{center}
\caption{
CEF parameters in the simulation of magnetic susceptibility and specific heat on CeNi$_{2}$Ge$_{2}$. 
Molecular field ($\lambda_a$) and constant susceptibility ($\chi_{0}$) along the $a$-axis are used for the simulation of magnetic susceptibility as tuning parameters in Eq.~\ref{eq:magIS}
}
\label{table: Magn}
\begin{tabular}{ccccc}
\\
\hline
$B_{2}^{0}$ (meV) & $B_{4}^{0}$ (meV) & $|B_{4}^{4}|$ (meV) &&\\
\hline
-0.99 & 0.066 & 0.36 &&\\
&&&&\\
\hline
$\lambda_a$ (mol/emu) & $\chi_{0}$ ($10^{-3}$ emu/mol)&&&\\
\hline
-100 & 1.2 &&&\\
\\
%0 ($c$) & 0 ($c$)&&&  \\ 
\hline
&&&&\\
\end{tabular}
\\ energy levels and wave functions \\
\begin{tabular}{ccccccc}
\hline
$E$ (K) & $\ket{+5/2}$ & $\ket{+3/2}$ & $\ket{+1/2}$ & $\ket{-1/2}$ & $\ket{-3/2}$ & $\ket{-5/2}$ \\
390 & 0 & 0 & 1 & 0 & 0 & 0 \\
390 & 0 & 0 & 0 & 1 & 0 & 0 \\
230 & 0 & $\sqrt{0.4}$ & 0 & 0 & 0 & $\sqrt{0.6}$ \\
230 & $\sqrt{0.6}$ & 0 & 0 & 0 & $\sqrt{0.4}$ & 0 \\
0 & 0 & $\sqrt{0.6}$ & 0 & 0 & 0 & $-\sqrt{0.4}$ \\
0 & $-\sqrt{0.4}$ & 0 & 0 & 0 & $\sqrt{0.6}$ & 0 \\ \hline
\end{tabular}
\end{center}
\end{table}
%-----------------------%-----------------------%-----------------------%-----------------------%-----------------------

%-------------------------%-------------------------%-------------------------%-------------------------%-------------------------%-------------------------%-------------------------%-------------------------%-------------------------%-------------------------


\begin{thebibliography}{99}
\bibitem{atomicCalcPar}
For the ion-model calculation, the atomic parameters of the $4f$-$4f$ and $3d$-$4f$ Coulomb interactions, the $3d$ and $4f$ spin-orbit couplings were obtained from the Hartree-Fock values for Ce$^{3+}$ ion~\cite{TholeXASLD}.
The reduction factor was set to 60\% (76\% and 98\%) for the $4f$-$4f$ interaction ($3d$-$4f$ interaction and $3d$ spin-orbit coupling) obtained by fitting our experimental isotropic XAS spectra at Ce $M_{4,5}$ edges.
\bibitem{WIEN2k_setup}
We used a $k$-point mesh of 46 x 46 x 46, and the cut off parameter $R_{MT}K_{max}$ was set to 7.0.
\end{thebibliography}
\end{document}